\documentstyle[aps,prd,preprint]{revtex}
\begin{document}
\tightenlines

\def\stacksymbols #1#2#3#4{\def\theguybelow{#2}
    \def\verticalposition{\lower#3pt}
    \def\spacingwithinsymbol{\baselineskip0pt\lineskip#4pt}
    \mathrel{\mathpalette\intermediary#1}}
\def\intermediary#1#2{\verticalposition\vbox{\spacingwithinsymbol
      \everycr={}\tabskip0pt
      \halign{$\mathsurround0pt#1\hfil##\hfil$\crcr#2\crcr
               \theguybelow\crcr}}}
\def\lapproxeq{\stacksymbols{<}{\sim}{2.5}{.2}}
\def\gapproxeq{\stacksymbols{>}{\sim}{3}{.5}}

\title{Unambiguous Probabilities in an Eternally Inflating Universe.}
\author{Alexander Vilenkin\footnote{Electronic address: vilenkin@cosmos2.phy.tufts.edu}}

\address{Institute of Cosmology,
        Department of Physics and Astronomy,\\
        Tufts University,
        Medford, Massachusetts 02155, USA}
\date{\today}
\maketitle

\begin{abstract}

``Constants of Nature'' and cosmological parameters may in fact be
variables related to some slowly-varying fields.  In models of eternal
inflation, such fields will take different values in different parts
of the universe.  Here I show how one can assign probabilities to
values of the ``constants'' measured by a typical observer.  This
method does not suffer from ambiguities previously discussed in the
literature. 

\end{abstract}

One of the striking aspects of the inflationary cosmology \cite{1} is
that, generically, inflation never ends \cite{2}. The evolution of the
inflaton field $\phi$ is influenced 
by quantum fluctuation, and as a result thermalization does not occur simultaneously in different parts of the universe. The dynamics of $\phi$ 
can be pictured as a random walk superimposed on the classical slow roll and can be described by a Fokker-Planck equation \cite{2,3,4}. On 
small scales, the fluctuations in the thermalization time give rise to a spectrum of small density fluctuations, but on large scales they make the 
universe extremely inhomogeneous. In most of the models one finds that at any time there are parts of the universe that are still inflating and that 
the total volume of the inflating regions is growing with time. This picture is often referred to as ``stochastic inflation'' or ``eternal inflation''.

Eternally inflating universes may contain thermalized regions characterized by different values of the constants of Nature and of the 
cosmological parameters. For example, the inflaton potential $V(\phi)$ may have several minima corresponding to different low-energy physics or to 
different values of the cosmological constant. A more interesting possibility is that the ``constants'' are related to some slowly-varying fields and 
take values in a continuous range. Examples are the effective gravitational constant in a Brans-Dicke-type theory \cite{5} and the density 
parameter $\Omega$ in some models of open inflation \cite{6}. Just like the inflaton field $\phi$, the fields $\chi_j (j = 1, ..., n)$ associated with 
the ``constants'' are subject to quantum fluctuations during inflation, and different regions of the universe thermalize with different values of 
$\chi_j$. An intriguing question is whether or not we can predict which values of the "constants" we are most likely to observe.

An eternally inflating universe is inhabited by numerous civilizations that will measure different values of $\chi_j$. (For simplicity, I do not 
distinguish between the fields $\chi_j$ and the associated "constants"). We can define the probability ${\cal P}(\chi)d^n\chi$ for $\chi_j$ to be in 
the intervals $d\chi_j$ as being proportional to the number of civilizations which will measure $\chi_j$ in that interval \cite{7}. Assuming that 
we are a typical civilization, we can expect to observe $\chi_j$ near the maximum of ${\cal P}(\chi)$ \cite{8}.

An immediate objection to this approach is that we are ignorant about the origin of life, let alone intelligence, and therefore the number of 
civilizations cannot be calculated. However, the approach can still be used to find the probability distribution for parameters which do not affect 
the physical processes involved in the evolution of life. The cosmological constant $\Lambda$, the density parameter $\Omega$ and the 
amplitude of density fluctuations $Q$ are examples of such parameters. We shall assume that our fields $\chi_j$ belong to this category. The 
probability for a civilization to evolve on a suitable planet is then independent of $\chi_j$, and instead of the number of civilizations we can use 
the number of habitable planets or, as a rough approximation, the number of galaxies.

If this philosophy is adopted, the problem of calculating ${\cal P}(\chi)$ can be split into two parts. The probability for us to observe certain 
values of $\chi_j$ is proportional to the volume of the regions where $\chi_j$ take specified values and to the density of galaxies in those regions. 
It is convenient to consider comoving regions and measure their volume at the time of thermalization. Then we can write
\begin{equation}
{\cal P}(\chi) \propto \nu(\chi){\cal P}_*(\chi).
\label{1}
\end{equation}
Here, ${\cal P}_*(\chi)d^n\chi$ is proportional to the volume of thermalized regions where $\chi_j$ take values in the intervals $d\chi_j$, and 
$\nu(\chi)$ is the number of galaxies that form per unit thermalized volume. The calculation of $\nu(\chi)$ is a standard astrophysical problem, 
which is completely unrelated to the calculation of the volume factor ${\cal P}_*(\chi)$. Our focus in this paper will be on the volume factor.

The problem one encounters in using Eq.(1) is that in an eternally
inflating universe the thermalized volume is infinite, even for a region of a finite comoving size. One can deal with this problem by simply 
introducing a time cutoff and including only regions that thermalized prior to some moment of time $t_c$. One finds, however, that the resulting 
probability distribution is extremely sensitive to the choice of the time variable $t$ \cite{4}. This gauge-dependence casts doubt on any 
conclusion reached using this approach.

An alternative procedure, suggested in \cite{9,VW}, is to introduce a $\chi$-dependent cutoff at the time $t_c(\chi)$, when all but a small 
fraction $\epsilon$ of the comoving volume destined to thermalize with $\chi_j$ in the intervals $d\chi_j$ has thermalized. The limit $\epsilon 
\to 0$ is taken after calculating the probabilities. It was shown in
\cite{9,VW}  that the resulting probability distribution is essentially insensitive to 
the choice of time parametrization. However, the same problem appears in a different guise. Linde and Mezhlumian \cite{10} have found a 
family of gauge-invariant cutoff procedures which includes the 
$\epsilon$-procedure described above. This indicates that the invariance requirement alone is not sufficient to define the probabilities uniquely.

Another source of uncertainty is associated with large values of the inflaton potential $V(\phi)$. Usually, 
$\phi$ is assumed to vary in an infinite range, with $V(\phi) \to \infty$ at $\phi \to \pm\infty$. Higher values of $V(\phi)$ correspond to higher 
rates of expansion, and thus the regions with Planckian and super-Planckian values of $V(\phi)$ expand at the 
highest rate. The effect of these regions on the dynamics of stochastic inflation is therefore very important \cite{11}. Since the physics at 
Planckian energy density is completely unknown, this results in a significant uncertainty \cite{4,VW}.

In the face of these difficulties, serious doubts have been expressed that a meaningful definition of probabilities in an eternally inflating universe 
is even in principle possible \cite{4,GBL,10}. However, these pessimistic conclusions may have been premature. In this paper, I am going to argue 
that unambiguous probabilities can in fact be defined in a wide class of models of eternal inflation.

The factor ${\cal P}_*(\chi)$ in Eq. (\ref{1}) is the probability distribution of the fields $\chi_j$ on the thermalization hypersurface $\Sigma_*$ 
which separates inflating and thermalized spacetime regions. It is a 3-dimensional spacelike surface which plays the role of the "big bang" for 
the thermalized regions.  In the 
case of several discrete vacua, $\Sigma_*$ consists of a number of disconnected pieces, each connected component corresponding to 
thermalization into a single vacuum \cite{LV}. Each connected piece of $\Sigma_*$ has an infinite volume. (The situation here is somewhat similar to 
open inflation, where thermalized regions inside the bubbles have the form of infinite, open Robertson-Walker universes). In order to determine 
the relative probability of different vacua, one has to compare the infinite volumes of the corresponding components of $\Sigma_*$, which is an 
inherently ambiguous task. This ambiguity is the source of the problems encountered in Refs. \cite{4,9,VW,10}.

Now, the key observation is that the situation may be greatly improved in the case of continuous fields $\chi_j$ varying in a finite range, $0 \leq 
\chi_j \leq  \Delta_j$. In this case, each connected part of $\Sigma_*$ is still infinite, but now different parts are not characterized by different 
values of $\chi_j$. On the contrary, $\chi_j$ run through all the range of their values on each connected part. Since the inflationary dynamics of 
the fields $\chi_j$ has a stochastic nature, the distributions of $\chi_j$ on different components of $\Sigma_*$  should be statistically equivalent. 
It is therefore sufficient to consider a single connected component. Moreover, since $\chi_j$ have a finite range, they will run through all of their 
values many times on any sufficiently large part of $\Sigma_*$. Hence, there is no need to deal with infinite hypersurfaces. The probability 
distribution ${\cal P}_*(\chi)$ can be determined by examining a large finite piece of $\Sigma_*$.

The above argument is an adaptation, to the case of stochastic inflation, of the argument given earlier by Garriga, Tanaka and Vilenkin \cite{12} 
for models of open inflation with a variable $\Omega$. In such models, each bubble contains regions with all possible values of $\Omega$, and 
we argued that the probability distribution for $\Omega$ can be determined by considering a single bubble.

The argument can also be extended to fields with an infinite range of variation, provided that the probability distributions of $\chi_j$ are 
concentrated within a finite range, with a negligible probability of finding $\chi_j$ very far away from that range.

Models with a finite range of $\chi_j$ are not difficult to construct. For example, $\chi_j$ could play the role of angular variables, with the 
inflaton field $\phi$ being the radial variable in the field space. Another attractive possibility is a $\sigma$-model-type theory in which both 
$\chi_j$ and $\phi$ take values on a compact manifold. In such models the potential is necessarily bounded and problems with super-Planckian 
densities are easily avoided.

The gauge-dependence of the probability distributions obtained using a constant-time cutoff can be understood as follows. Suppose that the time 
coordinate $t$ has been chosen to be the proper time, $t = \tau$, and that regions with $0 < \chi_1 < \Delta_1/2$ take a shorter time $\tau$ to 
thermalize than regions with $\Delta_1/2 < \chi_1 < \Delta_1$.  
Then, in the vicinity of the cutoff ($t = t_c$), the surface $\Sigma_*(t_c)$ will tend to include regions with $\chi_1 < \Delta_1/2$ and 
exclude regions with $\chi_1 > \Delta_1/2$. Suppose now that, despite the shorter roll-down time, regions with $\chi_1 < \Delta_1/2$ expand by 
a larger factor than regions with $\chi_1 > \Delta_1/2$. (This can be arranged with a suitable choice of the potential $V(\phi, \chi)$). 
Then, if we choose the scale factor to be the time coordinate, $t = a$, the surface $\Sigma_*(t_c)$ will tend to include regions with $\chi_1 > 
\Delta_1/2$ and exclude those with $\chi_1 < \Delta_1/2$. Thus, the cutoff at a fixed time $t_c$ is biased with respect to different types of 
regions, depending on the choice of the time coordinate. One might think that the probability distribution of $\chi$ on a hypersurface 
$\Sigma_*$ should be insensitive to variation of the boundary of $\Sigma_*$, provided that the volume of $\Sigma_*$ is sufficiently large. 
However, in the case of a fixed-time cutoff, $\Sigma_*$ is a multi-component hypersurface whose volume is dominated by small, newly formed 
thermalized regions. It is therefore no surprise that the resulting distribution ${\cal P}_*(\chi)$ is sensitive to the choice of the time variable.

Thus, what we need is an unbiasly selected portion of $\Sigma_*$. Such an unbiased cutoff is hard to implement in the Fokker-
Planck formalism \cite{2,3,4} which deals with probability distributions on equal-time surfaces. A more promising approach to the calculation of 
${\cal P}_*(\chi)$ is to use a numerical simulation of stochastic inflation. Simulations of this sort have been introduced in Ref. \cite{13} and 
have been further developed in  Ref. \cite{4}. 
In the present paper, I will not attempt to pursue this numerical approach. Instead, I am going to argue that in a wide class of 
models the probabilities can be calculated analytically.

Suppose the maximum of the potential $V(\phi, \chi)$ is at $\phi =
0$. We shall first consider models in which $V(\phi, \chi)$ is very
symmetric near 
 
the top, so that it is essentially independent of $\chi$ in the range $0 \leq \phi < \phi_1$, but may have significant $\chi$-dependence elsewhere. 
We shall assume that $\phi_1$ is in the slow-roll range, $\phi_q \ll  \phi_1 < \phi_*(\chi)$, where $\phi_q$ is the boundary between the quantum 
"diffusion" and deterministic slow roll regimes, and $\phi_*(\chi)$ is the value corresponding to the end of inflation. Let us consider a constant- 
$\phi$ hypersurface $\Sigma_0$: $\phi = \phi_0$ with $\phi_q \ll \phi_0 < \phi_1$. Since $\phi_0$ is in the slow-roll range, this hypersurface is 
(almost everywhere) spacelike \cite{9}. Moreover, all values of $\phi$ in the past of $\Sigma_0$ belong to the range $0 \leq \phi < \phi_0$ where the potential is 
independent of $\chi$. Therefore, we expect the distribution of $\chi_j$ on $\Sigma_0$ to be flat, ${\cal P}_0(\chi_j)$ = const, with all values of 
$\chi_j$ equally probable \cite{foot}.

The hypersurface $\Sigma_*$ is a spacelike hypersurface in the future of $\Sigma_0$. The probability distribution for $\chi_j$ on $\Sigma_*$ is 
related to that on $\Sigma_0$ by
\begin{equation}
{\cal P}_*(\chi) \propto {\cal P}_0(\chi^{(0)})\exp[3N(\chi^{(0)})]{\rm det}(\partial\chi^{(0)}/\partial\chi).
\label{2}
\end{equation}
Here, $\chi_j^{(0)}(\chi)$ are the values on $\Sigma_0$ that evolve, along the slow-roll trajectory, to the values $\chi_j$ on $\Sigma_*$, 
$N(\chi^{(0)})$ is the number of inflationary e-folding along that trajectory, $\exp(3N)$ is the volume expansion factor, and 
$\det(\partial\chi^{(0)}/\partial\chi)$ is the Jacobian transforming from variables $\chi_j^{(0)}$ to $\chi_j$. Now, ${\cal P}_0(\chi^{(0)}) = 
const$, and the functions $\chi^{(0)}(\chi)$ and $N(\chi^{(0)})$ can be straightforwardly determined from a given potential $V(\phi, \chi)$. 
Hence, Eq. (\ref{2}) solves the problem of finding the probability distribution for $\chi_j$ on $\Sigma_*$. This equation is significantly 
simplified in models where $\chi_j$ do not change much during the slow roll, so that $\chi_j^{(0)} \approx \chi_j$. In such models,
\begin{equation}
{\cal P}_*(\chi) \propto \exp[3N(\chi)].
\label{3}
\end{equation}

Eq.(2) disregards quantum fluctuations of $\chi_j$ and $\phi$ in the
slow-roll regime at $\phi\geq \phi_0$.  For $\chi_j$ this is
justified, provided that the average cumulative fluctuation is small
compared to $\Delta_j$.  This condition is not difficult to satisfy.
Small quantum fluctuations of $\phi$ are not correlated with $\chi_j$
and their effect vanishes to linear order in $\delta\phi$.  However,
quadratic and higher-order terms will give some corrections to
Eq.(2).  One may also be worried about the effect of very large
fluctuations which bring $\phi$ back to the quantum diffusion regime
\cite{4}. 
Although extremely rare, such fluctuations are likely to be present
for a sufficiently large piece of $\Sigma_0$.  If the probability of
large fluctuations were independent of $\chi_j$, then small patches of
$\Sigma_0$ which have such fluctuations in their future could be left
out, without any effect on the distribution (2).  If there is some
$\chi$-dependence, its effect will be to reduce ${\cal P}_*(\chi)$ for
those values of $\chi_j$ which give a higher probability of large
backwards jumps.  I expect this effect to be very small, but more work
is needed to obtain a quantitative estimate.

Large fluctuations back to the quantum regime may occur even after
thermalization \cite{GV}.  Then each thermalization surface $\Sigma_*$
may have an infinite number of other thermalization surfaces in its
future.  However, Eq.(2) can still be used to derive the probability
distribution for $\chi_j$ on any of these surfaces.
 
As an illustrative example, let us consider a two-field model with a potential
\begin{equation}
V(\phi, \chi) = V_0\cos^2(\alpha\phi)[1 + \lambda\sin^4(\alpha\phi)\cos(\beta\chi)].
\label{4}
\end{equation}
Here, $|\lambda| < 1$, $\alpha = \pi/\Delta_\phi$ and the range of $\phi$ is $0 \leq \phi \leq \Delta_\phi$, with the values $\phi = 0$ and $\phi = 
\Delta_\phi$ identified. Similarly, $\beta = \pi/\Delta_\chi$ with
$\chi = 0$ and $\chi = \Delta_\chi$ identified.  We shall assume that $\beta \lesssim \alpha \lesssim 1$, that is, 
$\Delta_\chi \gtrsim \Delta_\phi \gtrsim 1$.  (I use Planck units throughout the paper).
The top of the potential is at $\phi 
= 0$, and the true vacuum is at $\phi = \Delta_\phi/2$. 
  The field $\chi$ is massless in the true vacuum, while the mass of $\phi$ is $\chi$-dependent,
\begin{equation}
m_\phi^2(\chi) = 2V_0\alpha^2[1 + \lambda\cos(\beta\chi)].
\label{5}
\end{equation}

The boundary $\phi_q$ between the quantum diffusion and slow roll regimes is determined by the condition $|\partial H/\partial\phi| \sim H^2$, 
where $H(\phi, \chi) = [8\pi V(\phi, \chi)/3]^{1/2}$ is the inflationary expansion rate. We shall assume that $V_0 \ll \alpha^2$; then $\phi_q 
\sim V_0^{1/2}/\alpha^2 \ll \Delta_\phi$.
Since $\alpha \lesssim 1$, the problem of super-Planckian energies
does not arise in this case.

The potential (\ref{4}) is essentially independent of $\chi$ for $\phi \ll \Delta_\phi$, but the number of e-foldings $N$ and the amplitude of 
density fluctuations $\delta\rho/\rho$ are both $\chi$-dependent. For $\alpha \ll 1$, $\lambda \ll 1$ we find
\begin{equation}
N(\chi) \approx -4\pi\int_{\phi_0}^{\phi_*}{H\over{\partial H/\partial\phi}}d\phi \approx N_0 + {2\pi\lambda\over{\alpha^2}}\cos(\beta\chi)
\label{6}
\end{equation}
and \cite{1}
\begin{equation}
\delta\rho/\rho(\chi) \approx 200m_\phi(\chi).
\label{7}
\end{equation}
Here, $N_0$ is the number of e-foldings (from a specified reference point $\phi_0$) in the $\chi$-independent case ($\lambda = 0$), and I have 
used $\sin(\alpha\phi_*) \approx 1$, which is valid for $\alpha \lesssim 1$ (since $\phi_* \approx {\pi\over{2\alpha}} - {1\over{6}}$). 
The classical and quantum variations of $\chi$ during the relevant
part of the slow roll are, respectively, $\delta\chi_c/\Delta_\chi
\sim (\lambda/4\pi)(\beta/\alpha)^2 \ll 1$ and
$\delta\chi_q/\Delta_\chi\sim V_0^{1/2}\beta/\alpha\ll 1$, and can be 
neglected. This justifies the $\chi = const$ approximation in Eq. (\ref{6}) and the use of Eq. (\ref{3}) for the probability distribution,
\begin{equation}
{\cal P}_*(\chi) \propto \exp[6\pi\lambda\alpha^{-2}\cos(\beta\chi)].
\label{8}
\end{equation}
The corresponding distribution for $\delta\rho/\rho$ can easily be written using Eqs. (\ref{5}), (\ref{7}).

As a second example, we consider the spectrum of density perturbations in the standard model of inflation with a single field $\phi$. The 
perturbations are determined by quantum fluctuations $\delta\phi$ which can be regarded as random Gaussian variables with a dispersion 
$\sigma = H/2\pi$. A perturbation is produced on each comoving scale at the time when that scale crosses the horizon and has a gauge-invariant 
amplitude \cite{14}
\begin{equation}
\delta\rho/\rho \sim 8\pi H\delta\phi/H',
\label{9}
\end{equation}
where $H' = dH/d\phi$. Averaging over the Gaussian distribution gives the standard result $(\delta\rho/\rho)_{rms} \sim 4H^2/|H'|$.

Fluctuations of $\phi$ on different length scales are statistically independent and can be treated separately. We can therefore concentrate on a 
single scale corresponding to some value $\phi = \phi_0$, disregarding all the rest. The fluctuation $\delta\phi$ will take different values on 
different parts of the hypersurface $\Sigma_0: \phi = \phi_0$, so $\delta\phi$ plays the role of the field $\chi$ of our previous example. Linde {\it 
et. al.} \cite{15} have studied the probability distribution for the observed values of $\delta\phi$ using a constant proper time cutoff and arrived 
at a surprising conclusion that "typical" fluctuations can be much greater than the rms value $H/2\pi$. We shall see, however, that this large 
effect disappears in the present approach.

The probability distribution for $\delta\phi$ on $\Sigma_*$ can be found directly from Eq. (\ref{2}) with ${\cal P}_0(\delta\phi)$ being the 
initial Gaussian distribution on $\Sigma_0$. When the field $\phi$ undergoes a quantum jump $\delta\phi$, the number of e-foldings necessary 
to complete the classical rollover to $\phi_*$ is changed by $\delta N = 4\pi\delta\phi H_0/{H'}_0$. Hence, we obtain the distribution
\begin{equation}
{\cal P}_* \propto \exp\left[-{2\pi^2\over{H_0^2}}\left(\delta\phi - {3H_0^3\over{\pi {H'}_0}}\right)^2\right]
\label{10}
\end{equation}
which describes fluctuations with a nonzero mean value, $<\delta\phi> = 3H_0^3/\pi {H'}_0$. The inflaton tends to fluctuate in the direction 
opposite to the classical roll, because backward fluctuations prolong inflation and increase the volume factor. This effect, however, is 
hopelessly small: $<\delta\phi>/\sigma = 6H_0^2/{H'}_0 \sim (\delta\rho/\rho)_{rms} \ll 1$. We thus see that anthropic considerations do not 
substantially modify the standard calculations of the density
perturbation spectrum \cite{16}.

In conclusion, the approach described in this paper allows one to
assign unambiguous probabilities to continuously-varying ``constants''
measured by a typical observer.  In models where the inflaton
potential is nearly independent of the ``constants'' in the quantum
diffusion range, the probabilities can be calculated analytically.
Otherwise, a numerical simulation is needed to determine the
distribution ${\cal P}_0 (\chi^{(0)})$ in Eq. (\ref{2}). 

I am grateful to Jaume Garriga for useful comments on the manuscript.
This work was supported in part by the National Science Foundation.


\begin{thebibliography}{99}
\bibitem{foot}
If the fields $\chi_j$ are described by a $\sigma$-model with a
non-trivial metric in the $\chi$-space, so that the kinetic terms in
the Lagrangian are of the form
$K_{ij}(\chi)\partial_\mu\chi_i\partial^\mu\chi_j$, then ${\cal
P}_0(\chi) \propto [\det (K_{ij})]^{1/2}$.  In this paper I assume for
simplicity that $K_{ij} = (1/2)\delta_{ij}$.
\bibitem{16}
It should be noted that our results (\ref{3}) and (\ref{10}) are in agreement with those obtained using the $\epsilon$-prescription of 
Ref. \cite{9}. However, this agreement may not persist in the general
case.
\end{thebibliography}
\end{document}